\documentclass[12pt]{iopart}
\usepackage{cite}
\usepackage{graphicx}
\usepackage{lipsum} 

\begin{document}
\title[Quantum speed-up process of atom in dissipative cavity]{Quantum speed-up process of atom in dissipative cavity}
\author{Hong-Mei Zou\textsuperscript{1}\footnote{zhmzc1997@hunnu.edu.cn},Jianhe Yang\textsuperscript{1} , Danping Li\textsuperscript{2},Mao-Fa Fang\textsuperscript{1}\footnote{mffang@hunnu.edu.cn} }
\address{\textsuperscript{1} Synergetic Innovation Center for Quantum Effects and Applications, Key Laboratory of Low-dimensional Quantum Structures and Quantum Control of Ministry of Education, School of Physics and Electronics, Hunan Normal University, Changsha, 410081, P.R. China.\\
\textsuperscript{2} Faculty of Science, Guilin University of Aerospace Technology, Guilin 541004,
People’s Republic of China.}
\begin{abstract}
In this work, we obtain an analytical representation of the density operator of an atom in dissipative cavity when the reservoir is at zero temperature and the total number of excitation is N=1. We also investigated the quantum speed limit time(QSLT) of the atom and the non-Markovianity in the dynamics process. The results show that the QSLT and the non-Markovianity can be effectively manipulated by the atom-cavity coupling and the reservoir parameters. Both of the atom-cavity coupling and the detuning can induce a sudden transition from Markovian to non-Markovian dynamics and this transition is the main physical reason of the quantum speed-up process. The critical value of sudden transition from no speed-up to speed-up depends the atom-cavity coupling and the reservoir parameters. The corresponding physical explanation is also provided for our results.
\end{abstract}
\maketitle
\section{Introduction}
\hspace{12pt}

A fundamental and important task of quantum physics is to drive a given initial state to a target state at the maximal evolution speed, the so-called quantum speed limit(QSL) \cite{Anandan1,Vaidman1,Lloyd1,S.Luo1}. The minimal evolution time between two distinguishable states of a quantum system is defined as the quantum speed limit time(QSLT) \cite{Jones1,Zwierz1,Z.Y.Xu1,Xiangji1,Francesco1}, which sets a bound on the minimal time a quantum system needs to evolve between two distinguishable states. For a closed system, the QSLT is determined by the maximum of the Mandelstam-Tamm(MT) bound \cite{Mandelstam} and the Margolus-Levitin(ML) bound \cite{Margolus}, namely, $\tau=\max\{\frac{\pi\hbar}{2\Delta E},\frac{\pi\hbar}{2E}\}$, where $\Delta E$ and $E$ are the fluctuation and the mean value of the initial-state energy, respectively. For an open system, Deffner and Lutz obtained the unified bound of QSLT from the MT and ML types \cite{Deffner1}. S-X. Wu $et al.$ defines a new bound for the quantum speed limits and find that the non-Markovianity and the population of the excited state can fail to signal the quantum evolution acceleration, but the initial-state dependence is an important factor\cite{S.-X.Wu1}. The QSLT plays an important role in various areas of quantum physics such as quantum computation \cite{Lloyd2}, quantum communication \cite{Nielsen2}, quantum metrology \cite{Giovanetti1}, and quantum optimal control \cite{Hegerfeldt1,Avinadav1}.

For an open quantum system, due to the inevitable interaction with environment, the decoherence effect caused by the system-environment coupling would bring remarkable influence on the QSLT. In recent years, many efforts have been made in the study of QSLT of an open system \cite{Taddei1,delCampo1,Shao-xiong,Zhang1}. The authors in \cite{Liu} studied hybrid quantum-classical simulation of quantum speed limits in open quantum systems. As we known, the evolution process of an open system is Markovian in the weak system-environment coupling and the evolution process is non-Markovian in the strong system-environment coupling \cite{Davies}. This non-Markovian effect in the dynamics process may be described by the non-Markovianity \cite{Wolf1,Breuer1,Rivas1,S.Luo2,Zeng,He}. Deffner and Lutz \cite{Deffner1} found that, the non-Markovianity can speed up the quantum evolution process\cite{Wang1,Cianciaruso1,Ahansaz1}, that is, the non-Markovianity can reduce the QSLT below the actual evolution time however the QSLT can be equal to the actual evolution time in the Markovian process, which has been verified in different open sysytems \cite{YingJie,Z.Sun1,C.Liu1}. For example, the authors in Ref. \cite{YingJie} demonstrated that, under the weak system-environment coupling, the speed-up evolution of an open system can also be acquired by manipulating the strength of the classical field, that is, an external classical field can realize the transformation from Markovian to non-Markovian dynamics. However, the quantum speed-up process of an atom in dissipative cavity and the non-Markovianity in the dynamics process are also important and meaningful.

In this work, we obtain an analytical representation of the density operator of an atom in dissipative cavity when the cavity interacts with a Lorentzian and an Ohmic reservoirs at zero temperature, respectively. We also investigated the quantum speed limit time(QSLT) of the atom and the non-Markovianity in the dynamics process. The results show that, the atom-cavity coupling, the detuning and the stronger cavity-reservoir coupling can all accelerate the evolution of the atom and enhance the non-Markovianity in the dynamics process. The outline of the paper is the following. In Section 2, we introduce a physical model. In Section 3, we obtain the expressions of the quantum speed limit and the non-Markovianity. Results and discussions are provided in Section 4. Finally, we conclude with a brief summary of important results in Section 5.

\section{Model and Solution}
\hspace{12pt}
Our work is based on Jaynes-Cummings model \cite{Jaynes,Shore}. In previous theoretical and experimental studies, the JC model is usually regarded as an ideal model which means an atom in a perfect cavity. However, the actual cavity is often not perfect and there is a leakage of the field from the cavity, which is usually modeled by coupling of the cavity mode to the bosonic modes of the reservoir \cite{Scala0,Scala1}. Namely, we consider an atom(transition frequency $\omega_{0}$) in a single mode cavity interacting with a multi-mode reservoir. The dynamics of the total system is unitary and the Hamiltonian is given by
\begin{eqnarray}\label{EB301}
\hat{H}&=&\hat{H}_{a}+\hat{H}_{c}+\hat{H}_{e}+\hat{H}_{ac}+\hat{H}_{ce}.
\end{eqnarray}
where $\hat{H}_{a}$, $\hat{H}_{c}$ and $\hat{H}_{e}$ are the Hamiltonian of the atom, the single mode cavity and the multi-mode reservoir, respectively. $\hat{H}_{ac}$ and $\hat{H}_{ce}$ are the interaction Hamiltonian of the atom-cavity and the cavity-reservoir, respectively. They have the following forms
\begin{eqnarray}\label{EB302}
\hat{H}_{a}=\frac{1}{2}\omega_{0}\hat{\sigma}_{z}, \hat{H}_{c}=\omega_{0}\hat{a}^{\dag}\hat{a},  \hat{H}_{e}=\sum_{k}\omega_{k}\hat{c}_{k}^{\dag}\hat{c}_{k}
\end{eqnarray}
and
\begin{eqnarray}\label{EB303}
\hat{H}_{ac}=\Omega(\hat{a}\hat{\sigma}_{+}+\hat{a}^{\dag}\hat{\sigma}_{-}),
\hat{H}_{ce}=(\hat{a}+\hat{a}^{\dag})\sum_{k}g_{k}(\hat{c}_{k}^{\dag}+\hat{c}_{k})
\end{eqnarray}
In the above expressions, $\hat{\sigma}_{z}$ and $\hat{\sigma}_{\pm}$ are the Pauli matrices of the atom. $\hat{a}^{\dag}$($\hat{a}$) and $\hat{c}_{k}^{\dag}$($\hat{c}_{k}$) are the creation(annihilation) operators of the cavity and the k-th mode of reservoir with the frequency $\omega_{k}$, respectively. $\Omega$ and $g_{k}$ are the atom-cavity coupling and the cavity-reservoir coupling, respectively. For simplicity, we assume that the total system has only one initial excitation and the reservoir is at zero temperature in the following section. Performing the Born-Markov and the rotating wave approximations, tracing out the reservoir degree of freedom in the interaction picture and then going back to the Schr\"{o}dinger picture, we obtain the following master equation for the atom-cavity system in the dressed-state basis\{$|\alpha_{1,+}\rangle$, $|\alpha_{1,-}\rangle$, $|\alpha_{0}\rangle$\} \cite{Zou3} ($\hbar=1$)
\begin{eqnarray}\label{EB304}
\frac{d}{dt}\varrho(t)&=&-i[\hat{H}_{JC},\varrho(t)]+\frac{1}{2}\gamma_{1}(t)(\hat{b}_{1}^{-}\varrho(t)\hat{b}_{1}^{+}-\frac{1}{2}\{\hat{b}_{1}^{+}\hat{b}_{1}^{-},\varrho(t)\})\nonumber\\
&+&\frac{1}{2}\gamma_{2}(t)(\hat{b}_{2}^{-}\varrho(t)\hat{b}_{2}^{+}-\frac{1}{2}\{\hat{b}_{2}^{+}\hat{b}_{2}^{-},\varrho(t)\}),
\end{eqnarray}
where $\hat{H}_{JC}=\hat{H}_{a}+\hat{H}_{c}+\hat{H}_{ac}$. $\hat{b}_{1}^{+}=|\alpha_{1,-}\rangle\langle\alpha_{0}|$ and $\hat{b}_{1}^{-}=|\alpha_{0}\rangle\langle\alpha_{1,-}|$ express the jump operators between $|\alpha_{1,-}\rangle$ and $|\alpha_{0}\rangle$, and $\hat{b}_{2}^{+}=|\alpha_{1,+}\rangle\langle\alpha_{0}|$ and  $\hat{b}_{2}^{-}=|\alpha_{0}\rangle\langle\alpha_{1,+}|$ are the jump operators between $|\alpha_{1,+}\rangle$ and $|\alpha_{0}\rangle$, where $
|\alpha_{1,\pm}\rangle=\frac{1}{\sqrt{2}}(|1,g\rangle\pm|0,e\rangle)$ and $|\alpha_{0}\rangle=|0,g\rangle$. $\gamma_{1}(t)$ and $\gamma_{2}(t)$ are the time-dependent decay rates for $|\alpha_{1,-}\rangle$ and $|\alpha_{1,+}\rangle$, respectively.

We can acquire an analytical solution of the density operator $\varrho(t)$ from Eq.~(\ref{EB304}), then the density matrix $\rho(t)$ of the atom in the standard basis$\{|e\rangle, |g\rangle\}$ is also obtained by means of the representation transformation and taking a partial trace over the degree of freedom of the cavity. Suppose the initial state is $\{\rho_{11}(0),\rho_{10}(0),\rho_{01}(0),\rho_{00}(0)\}$, the density matrix $\rho(t)$ of the atom at all time $t$ is expressed as
\begin{eqnarray}\label{EB305}
\rho(t)=\left(
\begin{array}{cc}
|A(t)|^{2}\rho_{11}(0)& A(t)\rho_{10}(0)\\
A(t)^{\ast}\rho_{01}(0)&1-|A(t)|^{2}\rho_{11}(0)\\
\end{array}
\right),
\end{eqnarray}
where the probability amplitude $A(t)$ can be given by
\begin{eqnarray}\label{EB306}
A(t)&=&\frac{1}{2}\sum_{j=1}^{2}e^{-i\omega_{j}t}e^{-\frac{1}{4}\beta_{j}},
\end{eqnarray}
in which $\omega_{1}=\omega_{0}-\Omega$ $(\omega_{2}=\omega_{0}+\Omega)$ is the transition frequency of the dressed-states $|\alpha_{1,-}\rangle\rightarrow|\alpha_{0}\rangle$ ($|\alpha_{1,+}\rangle\rightarrow|\alpha_{0}\rangle$). The coefficients $
\beta_{j}=\int_{0}^{t}\gamma_{j}(t')dt'$ and $\gamma_{j}({t})=2\rm{Re}[{\int_{0}^{t}d\tau \int_{-\infty}^{+\infty}d\omega'e^{i(\omega_{j}-\omega')\tau}J(\omega')}]$ ($j=1,2$) when the spectral density of the reservoir is $J(\omega')$.

In view of Eq.~(\ref{EB305}), we can also write a time-local master equation \cite{Breuer} for the density operator $\rho(t)$ as
\begin{eqnarray}\label{EB309}
\frac{d}{dt}\rho(t)&=&\mathcal{L}\rho(t)\nonumber\\
&=&-\frac{i}{2}S(t)[\hat{\sigma}_{+}\hat{\sigma}_{-},\rho(t)]+\Gamma(t)\{\hat{\sigma}_{-}\rho(t)\hat{\sigma}_{+}\nonumber\\
&-&\frac{1}{2}\hat{\sigma}_{+}\hat{\sigma}_{-}\rho(t)-\frac{1}{2}\rho(t)\hat{\sigma}_{+}\hat{\sigma}_{-}\}
\end{eqnarray}
where $S(t)=-2\Im[\frac{\dot{A}(t)}{A(t)}]$ and $\Gamma(t)=-2\Re[\frac{\dot{A}(t)}{A(t)}]$ are the Lamb frequency shift and the decoherence rate of the atom, respectively. All the non-Markovian memory effects have been registered self-consistently in these time-dependent coefficients. $S(t)$ describes the contribution from the unitary part of the evolution under dynamical decoherence, and $\Gamma(t)$ characterizes the backflow of information from the environment to the atom when $\Gamma(t)$ is just negative.

\section{ Quantum speed limit and non-Markovianity}
The QSLT defines the bound of the minimal evolution time from an initial state $\rho(0)$ to a final state $\rho(\tau)$ by an actual evolution time $\tau$. When the initial state is $\rho(0)=|\varphi_{0}\rangle\langle\varphi_{0}|$ and its target state $\rho(\tau)$ is governed by the master equation $\dot{\rho}(t)=\mathcal{L}\rho(t)$ (see Eq.~(\ref{EB309})) with $\mathcal{L}$ being the positive generator of the dynamical semigroup, according to the unified lower bound derived by Deffner and Lutz, the QSLT is defined as $\tau_{QSL}=\sin^{2}\beta[\rho(0),\rho(\tau)]/\Lambda_{\tau}^{\infty}$, where $\beta[\rho(0),\rho(\tau)]= \arccos\sqrt{\langle\varphi_{0}|\rho_{\tau}|\varphi_{0}\rangle}$ indicates the Bures angle between $\rho(0)$ and $\rho(\tau)$, and $\Lambda_{\tau}^{\infty}=\tau^{-1}\int_{0}^{\tau}\|\mathcal{L}\rho(t)\|dt$ with the operator norm $\|B\|$ equaling to the largest eigenvalue of $\sqrt{B^{\dag}B}$ \cite{Deffner1}. If $\rho(0)=|e\rangle\langle e|$, we can obtain the QSLT of the atom from Eq.~(\ref{EB305}) as\cite{Z.Y.Xu1}
\begin{equation}\label{EB401}
\frac{\tau_{QSL}}{\tau}=\frac{1-|A(t)|^{2}}{\int_{0}^{\tau}\partial_{t}|A(t)|^{2}dt},
\end{equation}
where $|A(t)|^{2}$ is the population of the excited state $|e\rangle$ at time $t$ and given by Eq.~(\ref{EB306}).

The non-Markovianity in the dynamics process (from $\rho(0)$ to $\rho(\tau)$) can be calculated by $\mathcal{N}=\max_{\rho_{1,2}}\int_{\sigma>0}\sigma(t,\rho_{1,2}(0))dt$ \cite{Z.Y.Xu1,Deffner1,HaiBin}, in which $\sigma(t,\rho_{1,2}(0))=\frac{d}{dt}\mathcal{D}(\rho_{1}(t),\rho_{2}(t))$ is the time rate of change of the trace distance. The trace distance is defined as $\mathcal{D}(\rho_{1}(t),\rho_{2}(t))=\frac{1}{2}\rm{Tr}\|\rho_{1}(t)-\rho_{2}(t)\|$ and indicates the distinguishability between the two states $\rho_{1,2}(t)$ evolving from their respective initial forms $\rho_{1,2}(0)$ \cite{Breuer1}. If $\sigma(t,\rho_{1,2}(0))>0$, $\mathcal{D}(\rho_{1}(t),\rho_{2}(t))$ increases with time increasing because the information flows back from the environment to the system, there is $\mathcal{N}>0$, the dynamics process of the system is non-Markovian. When $\sigma(t,\rho_{1,2}(0))<0$, there is $\mathcal{N}=0$, the dynamics process of the system is Markovian because the information flows irreversibly from the system to the environment. Thus the non-Markovianity describes the total backflow of quantum information between the system and the environment. For the atom in Eq.~(\ref{EB305}), it can been proven that the optimal pair of initial states to maximize $\mathcal{N}$ are $\rho_{1}(0)=|e\rangle\langle e|$ and $\rho_{2}(0)=|g\rangle\langle g|$. We can get $\mathcal{D}(\rho_{1}(t),\rho_{2}(t))=|A(t)|^{2}$ by using Eq.~(\ref{EB305}), therefor the non-Markovianity is written as
\begin{equation}\label{EB402}
\mathcal{N}=\frac{1}{2}[\int_{0}^{\tau}|\partial_{t}|A(t)|^{2}|dt+|A(\tau)|^{2}-1].
\end{equation}

From Eqs.~(\ref{EB401})-~(\ref{EB402}), the relationship \cite{Z.Y.Xu1} between the QSLT and the non-Markovianity can be obtained as
\begin{equation}\label{EB403}
\frac{\tau_{QSL}}{\tau}=\frac{1-|A(\tau)|^{2}}{1-|A(\tau)|^{2}+2\mathcal{N}},
\end{equation}
Eq.~(\ref{EB403}) shows that the QSLT is equal to the actual evolution time because the information flows irreversibly from the atom to the environment when $\mathcal{N}=0$, but the QSLT is smaller than the actual evolution time because the information flows back from the environment to the atom when $\mathcal{N}>0$. That is, the non-Markovianity in the dynamics process can lead to faster quantum evolution and smaller QSLT.

\section{ Result and discusion}
\subsection{ Lorentzian spectral density}
We assume that the spectral density of the reservoir is $
J(\omega')=\frac{1}{2\pi}\frac{\gamma_{0}\lambda^{2}}{(\omega_{0}-\omega'-\delta)^{2}+\lambda^{2}}$, where $\delta$ is the detuning between the frequency of the cavity mode and the center frequency of spectrum. The parameter $\lambda$ defines the spectral width of coupling, which is connected to the reservoir correlation time $\tau_{R}$ by $\tau_{R}$=$\lambda^{-1}$ and the parameter $\gamma_{0}$ is related to the relaxation time scale $\tau_{S}$ by $\tau_{S}$=$\gamma_{0}^{-1}$. If $\lambda>2\gamma_{0}$, the relaxation time of the cavity is greater than the reservoir correlation time and the dynamical evolution of the system is essentially Markovian, which means a weak cavity-reservoir coupling regime. For $\lambda<2\gamma_{0}$, the reservoir correlation time is greater than or of the same order as the relaxation time and non-Markovian effects become relevant, which is a strong cavity-reservoir coupling regime \cite{Zou1,Bellomo1}. We can obtain $\gamma_{j}({t})=\frac{\gamma_{0}\lambda^{2}}{(\omega_{0}-\omega_{j}-\delta)^{2}+\lambda^{2}}
\{1+(\frac{\omega_{0}-\omega_{j}-\delta}{\lambda}\sin((\omega_{0}-\omega_{j}-\delta)t)
-\cos((\omega_{0}-\omega_{j}-\delta)t))e^{-\lambda t}\}$, then $\beta_{j}$ in Eq.~(\ref{EB306}) is stated as
\begin{eqnarray}\label{EB313}
\beta_{1}&=&\frac{\gamma_{0}\lambda^{2}}{(\Omega-\delta)^{2}+\lambda^{2}}[t-\frac{2(\Omega-\delta)e^{-\lambda t} \sin((\Omega-\delta)t)}{(\Omega-\delta)^{2}+\lambda^{2}}\nonumber\\
&+&\frac{(\lambda^{2}-(\Omega-\delta)^{2})(e^{-\lambda t}\cos((\Omega-\delta)t)-1)}{\lambda((\Omega-\delta)^{2}+\lambda^{2})}],\nonumber\\
\beta_{2}&=&\frac{\gamma_{0}\lambda^{2}}{(\Omega+\delta)^{2}+\lambda^{2}}[t-\frac{2(\Omega+\delta)e^{-\lambda t} \sin((\Omega+\delta)t)}{(\Omega+\delta)^{2}+\lambda^{2}}\nonumber\\
&+&\frac{(\lambda^{2}-(\Omega+\delta)^{2})(e^{-\lambda t}\cos((\Omega+\delta) t)-1)}{\lambda((\Omega+\delta)^{2}+\lambda^{2})}].
\end{eqnarray}

In Figure 1, we depict the curves of the QSLT and the non-Markovianity as a function of the atom-cavity coupling $\Omega$ in the resonance($\delta=0$). Figure 1(a) gives the non-Markovianity as a function of $\Omega$ in the weak ($\lambda=5\gamma_{0}$) and strong ($\lambda=0.01\gamma_{0}$) cavity-reservoir coupling regimes, respectively. The results show that, there is a same critical value $\Omega_{c}$ that $\mathcal{N}$ steeply increases from 0 for $\lambda=5\gamma_{0}$ and $\lambda=0.01\gamma_{0}$, and $\mathcal{N}$ is larger in the latter than in the former when $\Omega>\Omega_{c}$. Figure 1(b) exhibits the QSLT as a function of $\Omega$ in the weak($\lambda=5\gamma_{0}$) and strong($\lambda=0.01\gamma_{0}$) cavity-reservoir coupling regimes, respectively. From Figure 1(b), we find that a sudden transition from no speed-up to speed-up also occurs at a same critical point $\Omega_{c}$ in both of $\lambda=5\gamma_{0}$ and $\lambda=0.01\gamma_{0}$, and $\frac{\tau_{QSL}}{\tau}$ is smaller in the latter than in the former when $\Omega>\Omega_{c}$. In order to show more clearly the dependency relationship of the QSLT and the non-Markovianity, we draw their curves when $\lambda=5\gamma_{0}$ in Figure 1(c). It is interesting to find that both $\tau_{QSL}$ and $\mathcal{N}$ have a same critical point $\Omega_{c}=1.58\gamma_{0}$, which the atom presents a dramatic non-Markovian effect and speed-up process. That is to say, $\mathcal{N}$ remains zero and $\frac{\tau_{QSL}}{\tau}$ stays at 1 when $\Omega<\Omega_{c}$, but $\mathcal{N}$ will increase at different rates and $\frac{\tau_{QSL}}{\tau}$ experiences a sudden transition from no speed-up to speed-up evolution and then decreases periodically when $\Omega\geq\Omega_{c}$. Therefore we obtain the interesting result that the atom-cavity coupling and the smaller $\lambda$ value can all enhance the non-Markovianity in the dynamics process and speed up the evolution of the atom.

\begin{center}
\includegraphics[width=5cm,height=4cm]{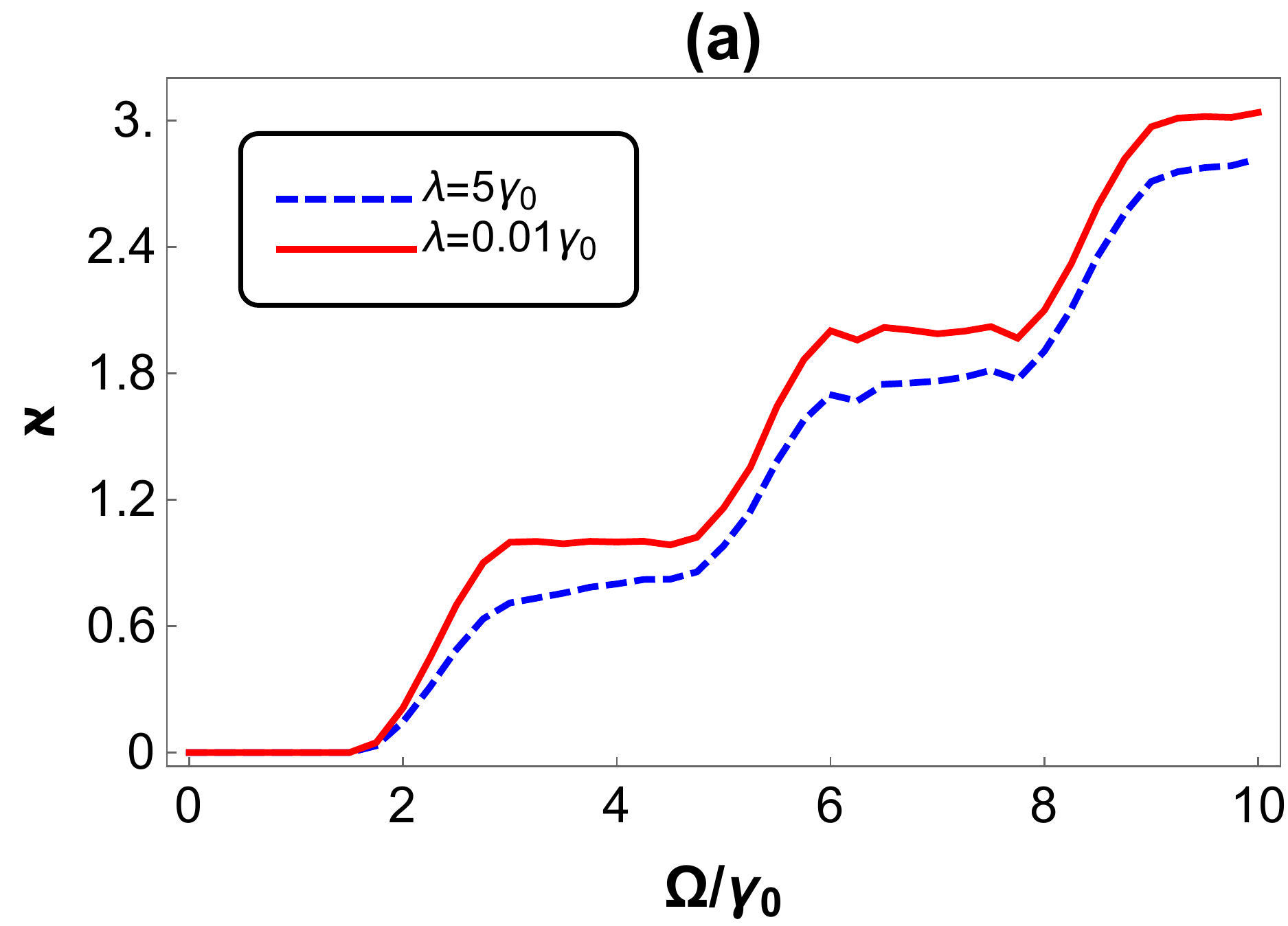}
\includegraphics[width=5cm,height=4cm]{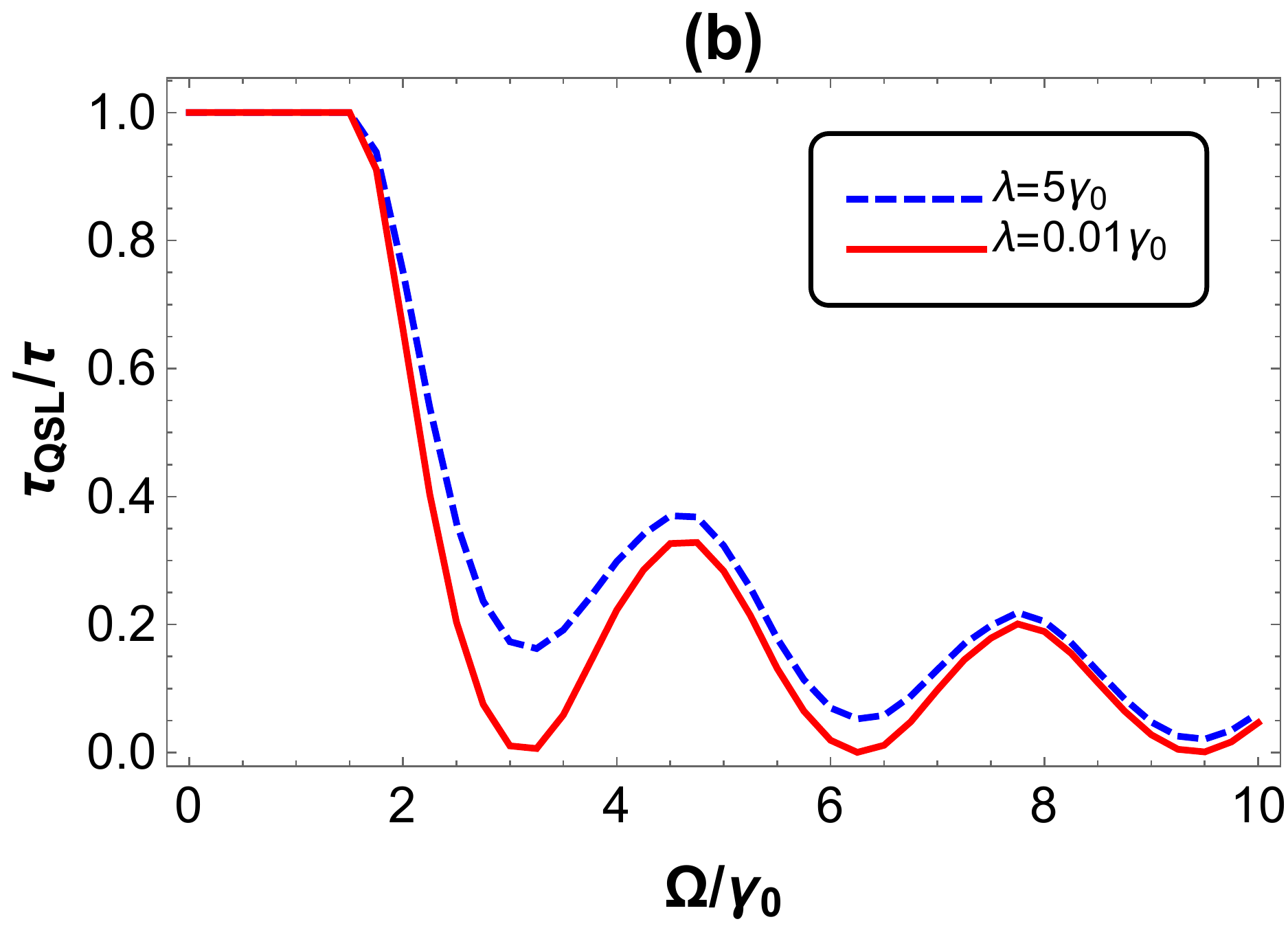}
\includegraphics[width=5.3cm,height=4cm]{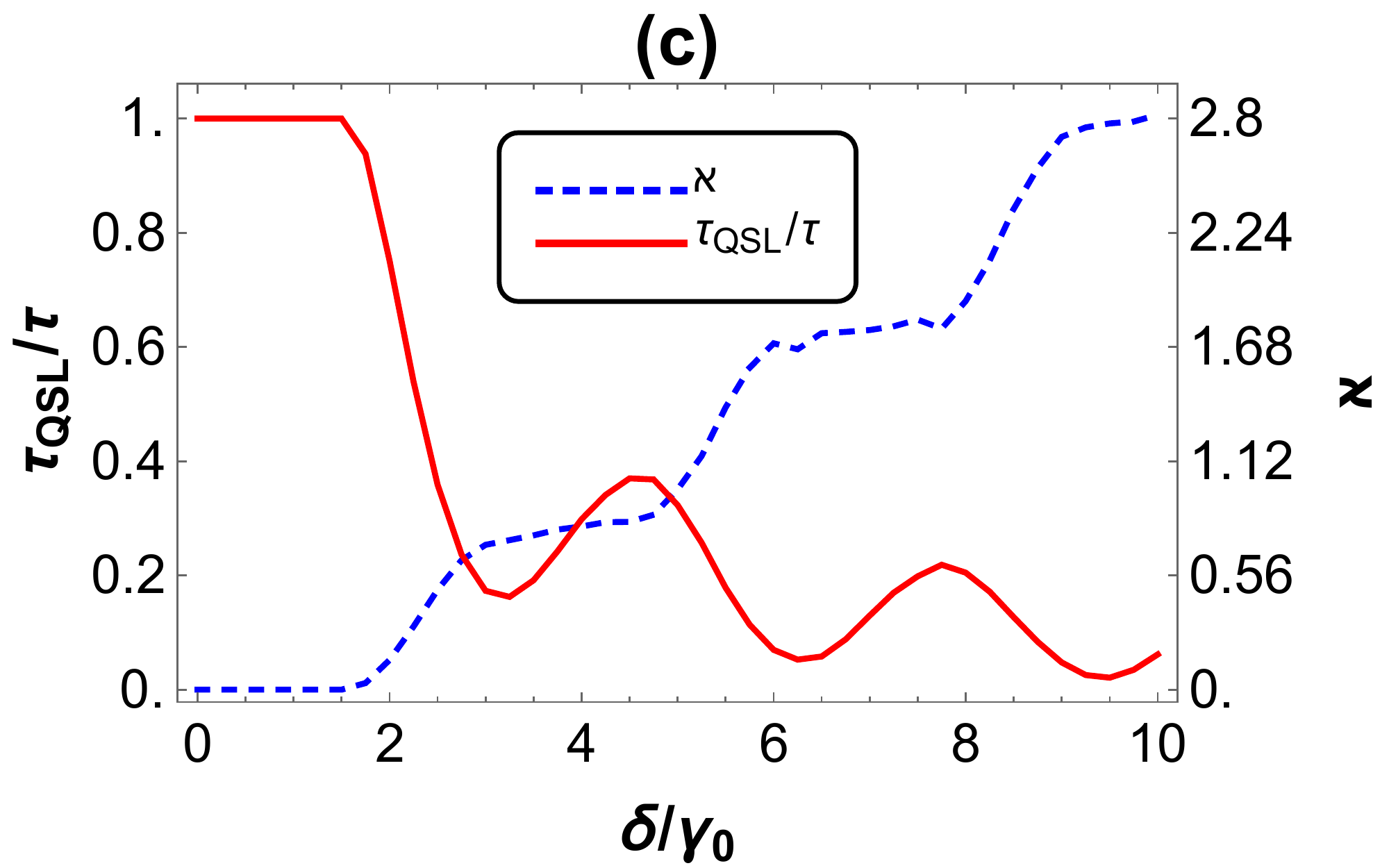}
	\parbox{14.5cm}{\small{\bf Figure 1.}
		(Color online)Influences of the atom-cavity coupling on the QSLT and the non-Markovianity when $\delta=0$. (a)non-Markovianity as a function of the coupling strength $\Omega$ in the weak-coupling($\lambda=5\gamma_{0}$, blue dashing line) and strong-coupling($\lambda=0.01\gamma_{0}$, red solid line) regimes, respectively; (b)QSLT as a function of the coupling strength $\Omega$ in the weak-coupling($\lambda=5\gamma_{0}$, blue dashing line) and strong-coupling($\lambda=0.01\gamma_{0}$, red solid line) regimes, respectively; (c)QSLT(red solid line) and non-Markovianity(blue dashing line) as a function of the coupling strength $\Omega$ when $\lambda=5\gamma_{0}$.}
\end{center}

In Figure 2, we describe the influence of the detuning on the QSLT and the non-Markovianity when $\Omega=0.01\gamma_{0}$. Figure 2(a) gives the non-Markovianity as a function of the detuning in the weak ($\lambda=5\gamma_{0},3\gamma_{0}$) and strong ($\lambda=\gamma_{0},0.5\gamma_{0}$) cavity-reservoir coupling regimes, respectively. The results show that, there are different critical values that $\mathcal{N}$ steeply increases from 0, for instances, $\delta_{c}=18.3\gamma_{0}$ when $\lambda=5\gamma_{0}$, but $\delta_{c}=3.5\gamma_{0}$ when $\lambda=0.5\gamma_{0}$. Figure 2(b) exhibits the QSLT as a function of the detuning in the weak ($\lambda=5\gamma_{0},3\gamma_{0}$) and strong ($\lambda=\gamma_{0},0.5\gamma_{0}$) cavity-reservoir coupling regimes, respectively. From Figure 2(b), we find that the sudden transition from no speed-up to speed-up also occurs at different critical points, for examples, $\delta_{c}=18.3\gamma_{0}$($\lambda=5\gamma_{0}$) and $\delta_{c}=3.5\gamma_{0}$($\lambda=0.5\gamma_{0}$), and $\frac{\tau_{QSL}}{\tau}$ decreases monotonically with increasing $\delta$ when $\lambda>2\gamma_{0}$ but decreases periodically with increasing $\delta$ when $\lambda<2\gamma_{0}$.  Figure 2(c) shows that the QSLT and the non-Markovianity as a function of the detuning when $\lambda=3\gamma_{0}$ and $\Omega=0.01\gamma_{0}$. We see that a critical point of sudden transitions of both $\tau_{QSL}$ and $\mathcal{N}$ is at $\delta_{c}=11.0\gamma_{0}$. $\mathcal{N}$ remains zero and $\frac{\tau_{QSL}}{\tau}$ stays at 1 when $\delta<\delta_{c}$, however $\mathcal{N}$ experiences a sudden increase and $\frac{\tau_{QSL}}{\tau}$ experiences a sudden decrease when $\delta>\delta_{c}$. That is to say, the detuning can also enhance the non-Markovianity in the dynamics process and speed up the evolution of the atom. The larger $\lambda$ value corresponds to a greater critical value when $\Omega=0.01\gamma_{0}$.

\begin{center}
	\includegraphics[width=5cm,height=4cm]{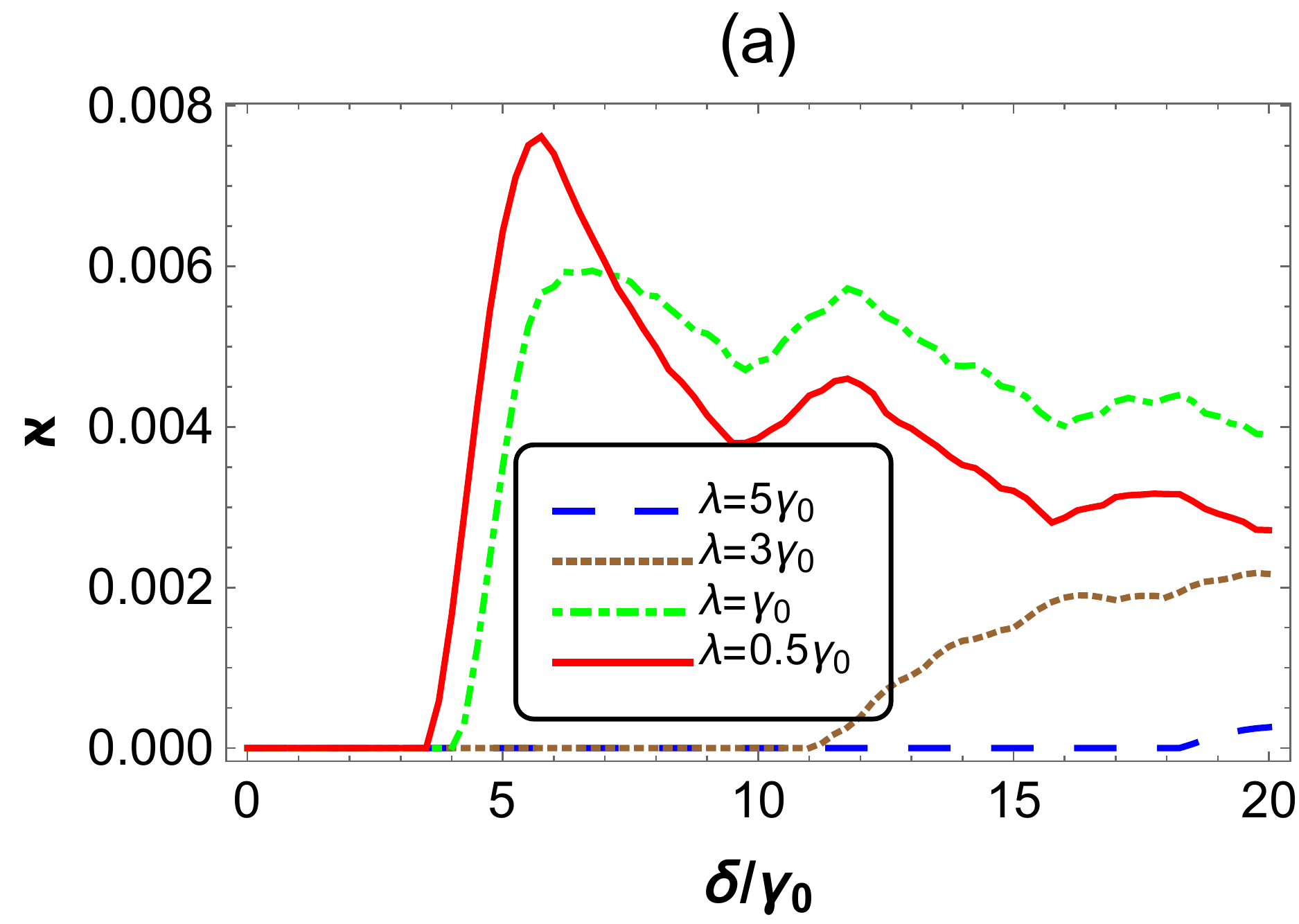}
	\includegraphics[width=5cm,height=4cm]{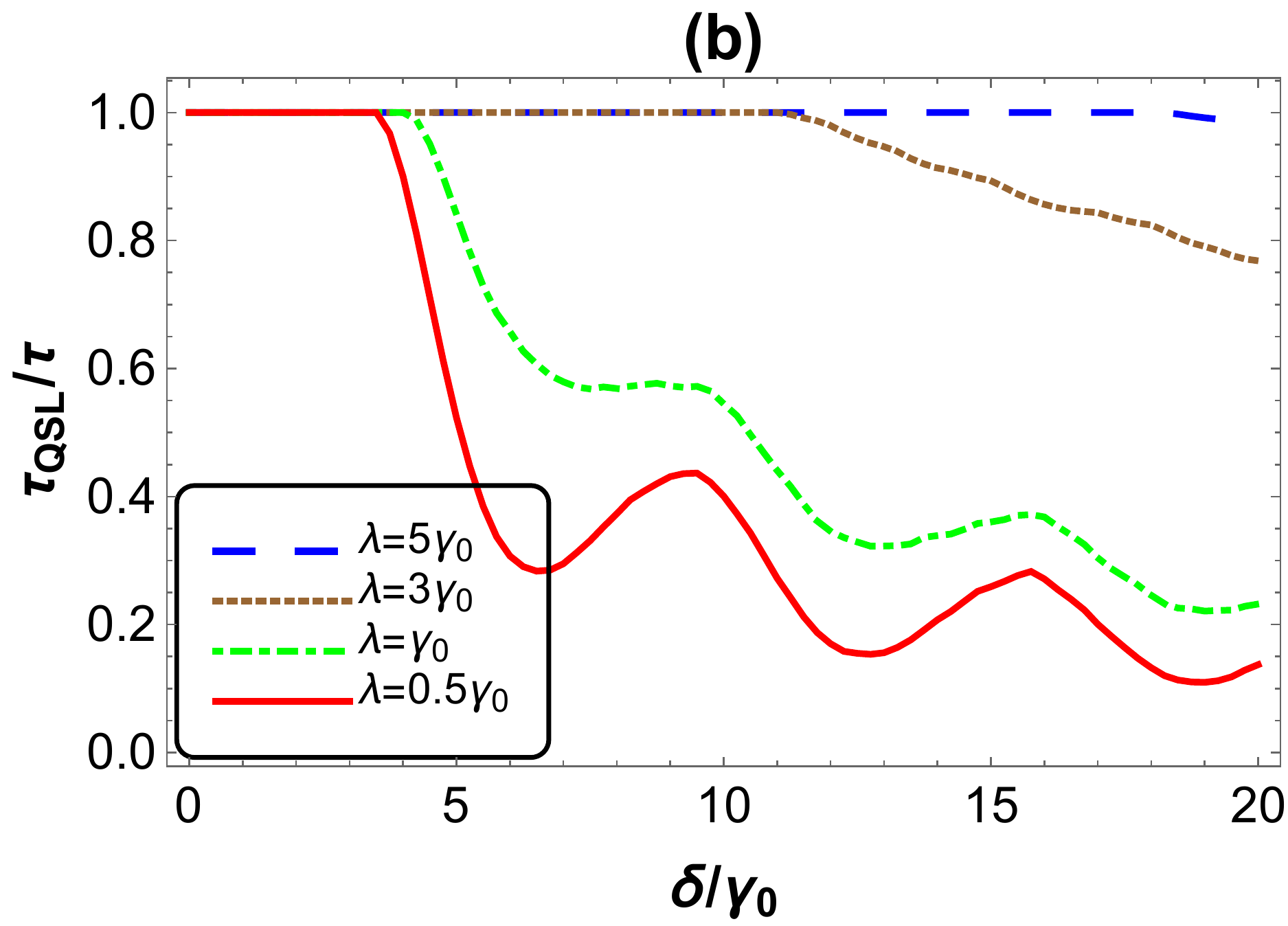}
	\includegraphics[width=5.3cm,height=4cm]{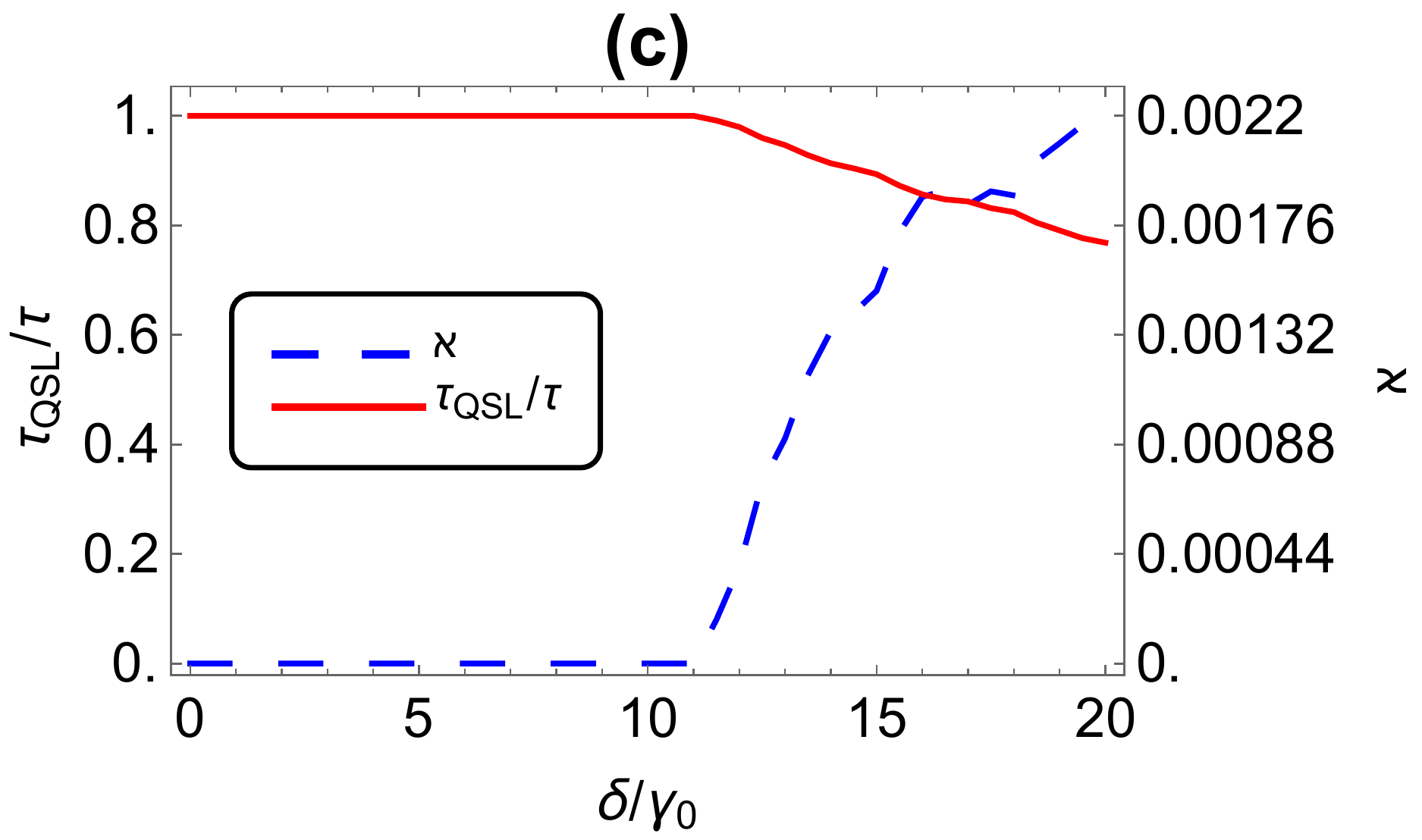}
	\parbox{14.5cm}{\small{\bf Figure 2.}
		(Color online)Influences of the detuning on QSLT and non-Markovianity when $\Omega=0.01\gamma_{0}$. (a)non-Markovianity as a function of the detuning $\delta$ in the weak-coupling($\lambda=5\gamma_{0}$, blue dashing line; $\lambda=3\gamma_{0}$, brown dotted line) and strong-coupling($\lambda=\gamma_{0}$, green dot-dashed line; $\lambda=0.5\gamma_{0}$, red solid line) regimes when $\Omega=0.01\gamma_{0}$, respectively; (b)QSLT as a function of the detuning $\delta$ in the weak-coupling($\lambda=5\gamma_{0}$, blue dashing line; $\lambda=3\gamma_{0}$, brown dotted line) and strong-coupling($\lambda=\gamma_{0}$, green dot-dashed line; $\lambda=0.5\gamma_{0}$, red solid line) regimes when $\Omega=0.01\gamma_{0}$, respectively; (c)QSLT(red solid line) and non-Markovianity(blue dashing line) as a function of the detuning $\delta$ when $\lambda=3\gamma_{0}$ and $\Omega=0.01\gamma_{0}$.}
\end{center}

From Figure 1 and Figure 2, we found that both of the atom-cavity coupling and the detuning can induce the transition from Markovian to non-Markovian dynamics and this transition is the main physical reason of the speed-up process. The sudden transition from no speed-up to speed-up occurs at a same critical point for different $\lambda$ when $\delta=0$. However, when $\Omega=0.01\gamma_{0}$, the sudden transitions can occur at different critical points for different $\lambda$. The smaller $\lambda$ value causes to the quicker speedup evolution.

We may give the physical interpretation of the above results. In the total system, the quantum information is exchanged between the atom with the cavity and between the cavity with the reservoir. For the atom we're interested in, both the cavity and its outside reservoir are regarded as the atomic environment. The non-Markovianity $\mathcal{N}$ is used to quantify the effect of all environment parameters (including the coupling $\Omega$, the detaining $\delta$, the spectral width $\lambda$ and the dissipation rate $\gamma_{0}$) on the atomic dynamical behavior, namely, the non-Markovianity $\mathcal{N}$ is the collective expression of the several parameters, as shown in Eqs.~(\ref{EB306}), ~(\ref{EB402}) and ~(\ref{EB313}). The larger the atom-cavity coupling $\Omega$, the more information the cavity feeds back to the atom, so the non-Markovianity $\mathcal{N}$ will increase with increasing $\Omega$. When the detuning $\delta=0$, the information flows irreversibly from the cavity to the reservoir due to the dissipation of reservoir if $\lambda>2\gamma_{0}$ but the information will flow back from the reservoir to the cavity due to the memory and feedback effect of reservoir if $\lambda<2\gamma_{0}$, therefore the smaller $\lambda$ value causes to the larger non-Markovianity $\mathcal{N}$. On the other hand, the influence of the cavity on the atom is obviously greater than that of the reservoir on the atom, the critical value of sudden transition is determined by $\Omega$ rather than $\lambda$, as shown in Figure 1(a). From Eq.~(\ref{EB313}), we can find that, the physical principle that the detuning can accelerate the information exchange between the cavity and the reservoir is similar to that the atom-cavity coupling can accelerate the information exchange between the atom and the cavity. The critical value of sudden transition decreases with increasing $\delta$ when $\Omega$ is very small ( $\Omega=0.01\gamma_{0}$), as shown in Figure 2(a). The influence of all parameters on the QSLT can be similarly analyzed, as shown in Figure 1(b) and 2(b), but we omit it in order to save the space. Besides, when $\mathcal{N}=0$, the information flows irreversibly from the atom to the environment so that the atom evolves at the actual speed and the QSLT is equal to the actual evolution time. When $\mathcal{N}>0$, the information flows back from the environment to the atom thus the atom evolution is accelerated and the QSLT is smaller than the actual evolution time. The larger $\mathcal{N}$ induces the smaller QSLT, as shown in Figure 1(c) and 2(c), which is according to Eq.~(\ref{EB403}). 

\subsection{ Ohmic spectral density}
We suppose that the reservoir has the Ohmic spectral density with a Lorentz-Drude cutoff function $J(\omega')=\frac{2\omega'}{\pi}\frac{\omega_{c}^{2}}{\omega_{c}^{2}+\omega'^{2}}$, where $\omega'$ is the frequency of the reservoir and $\omega_{c}$ is the cut-off frequency which depends on the reservoir coupling strength. The case $\omega_{c}\ll\omega_{0}$ implies an essentially strong cavity-reservoir coupling regime, while the case $\omega_{c}\gg\omega_{0}$ indicates a weak cavity-reservoir coupling regime \cite{Sinayskiy,Zou2,Eckel,CuiW}.
The decay rates is $\gamma_{j}({t})=\frac{4\omega_{c}^2}{\omega_{j}^2+\omega_{c}^2}[\omega_{j}(1-e^{-w_{c}t}\cos(\omega_{j} t))-\omega_{c}e^{-w_{c}t}\sin(\omega_{j}t)]$ and
$\beta_{j}$ in Eq.~(\ref{EB306}) has the following forms
\begin{eqnarray}\label{EB323}
\beta_{1}&=&\frac{4\omega_{c}^2}{[(\omega_{0}-\Omega)^2+\omega_{c}^2]^2}\{[(\omega_{0}-\Omega)^2
+\omega_{c}^2](\omega_{0}-\Omega) t\nonumber\\
&+&2\omega_{c}(\omega_{0}-\Omega)[e^{-w_{c}t}\cos((\omega_{0}-\Omega)t)-1]\nonumber\\
&-&[(\omega_{0}-\Omega)^2-\omega_{c}^2]e^{-w_{c}t}\sin((\omega_{0}-\Omega)t)\},\nonumber\\
\beta_{2}&=&\frac{4\omega_{c}^2}{[(\omega_{0}+\Omega)^2+\omega_{c}^2]^2}\{[(\omega_{0}+\Omega)^2
+\omega_{c}^2](\omega_{0}+\Omega) t\nonumber\\
&+&2\omega_{c}(\omega_{0}+\Omega)[e^{-w_{c}t}\cos((\omega_{0}+\Omega)t)-1]\nonumber\\
&-&[(\omega_{0}+\Omega)^2-\omega_{c}^2]e^{-w_{c}t}\sin((\omega_{0}+\Omega)t)\}.
\end{eqnarray}

In Figure 3, we display the curves of the QSLT and the non-Markovianity with the coupling $\Omega$ increasing. Figure 3(a) exhibits the non-Markovianity as a function of $\Omega$ in the weak ($\frac{\omega_{c}}{\omega_{0}}=10,2$) and strong ($\frac{\omega_{c}}{\omega_{0}}=0.3,0.1$) cavity-reservoir coupling regimes, respectively. The results show that, there are different critical values that $\mathcal{N}$ steeply increases from 0 when $\frac{\omega_{c}}{\omega_{0}}>1$, i.e. $\Omega_{c}=0.90\omega_{0}$ when $\frac{\omega_{c}}{\omega_{0}}=10$, but $\Omega_{c}=0.66\omega_{0}$ when $\frac{\omega_{c}}{\omega_{0}}=2$, and $\mathcal{N}$ increases monotonically with $\Omega$. However, there is a same critical value $\Omega_{c}=0.18\omega_{0}$ when $\frac{\omega_{c}}{\omega_{0}}<1$, and $\mathcal{N}$ obviously enlarges with decreasing $\frac{\omega_{c}}{\omega_{0}}$. Figure 3(b) indicates the QSLT as a function of $\Omega$ in the weak ($\frac{\omega_{c}}{\omega_{0}}=10,2$) and strong ($\frac{\omega_{c}}{\omega_{0}}=0.3,0.1$) cavity-reservoir coupling regimes, respectively. From Figure 3(b), we find that the sudden transition from no speed-up to speed-up occurs at different critical points $\Omega_{c}=0.90\omega_{0}$(when $\frac{\omega_{c}}{\omega_{0}}=10$) and $\Omega_{c}=0.66\omega_{0}$(when $\frac{\omega_{c}}{\omega_{0}}=2$) if $\frac{\omega_{c}}{\omega_{0}}>1$ while the sudden transition occurs at a same critical point $\Omega_{c}=0.18\omega_{0}$ when $\frac{\omega_{c}}{\omega_{0}}<1$. Besides, $\frac{\tau_{QSL}}{\tau}$ decreases monotonically with increasing $\Omega$ if $\frac{\omega_{c}}{\omega_{0}}>1$ while decreases periodically with increasing $\Omega$ if $\frac{\omega_{c}}{\omega_{0}}<1$. Figure 3(c) shows the dependency relationship of the QSLT and the non-Markovianity as a function of $\Omega$ when $\frac{\omega_{c}}{\omega_{0}}=2$. A same critical point for both $\tau_{QSL}$ and $\mathcal{N}$ is at  $\Omega_{c}=0.66\omega_{0}$. $\mathcal{N}$ remains zero and $\frac{\tau_{QSL}}{\tau}$ stays at 1 when $\Omega<\Omega_{c}$ but $\mathcal{N}$ will increases monotonically and $\frac{\tau_{QSL}}{\tau}$ decreases monotonically when $\Omega>\Omega_{c}$. Therefore the atom-cavity coupling and the smaller $\frac{\omega_{c}}{\omega_{0}}$ value can all enhance the non-Markovianity and speed up the evolution process, i.e., they are the main physical reason of the transition from Markovian to non-Markovian dynamics and the speed-up process. This sudden transitions can occur at different critical points  when $\frac{\omega_{c}}{\omega_{0}}>1$. However, 
at $\frac{\omega_{c}}{\omega_{0}}<1$, 
the sudden transition from no speed-up to speed-up occurs at a same critical point when $\frac{\omega_{c}}{\omega_{0}}<1$. 

\begin{center}
	\includegraphics[width=5cm,height=4cm]{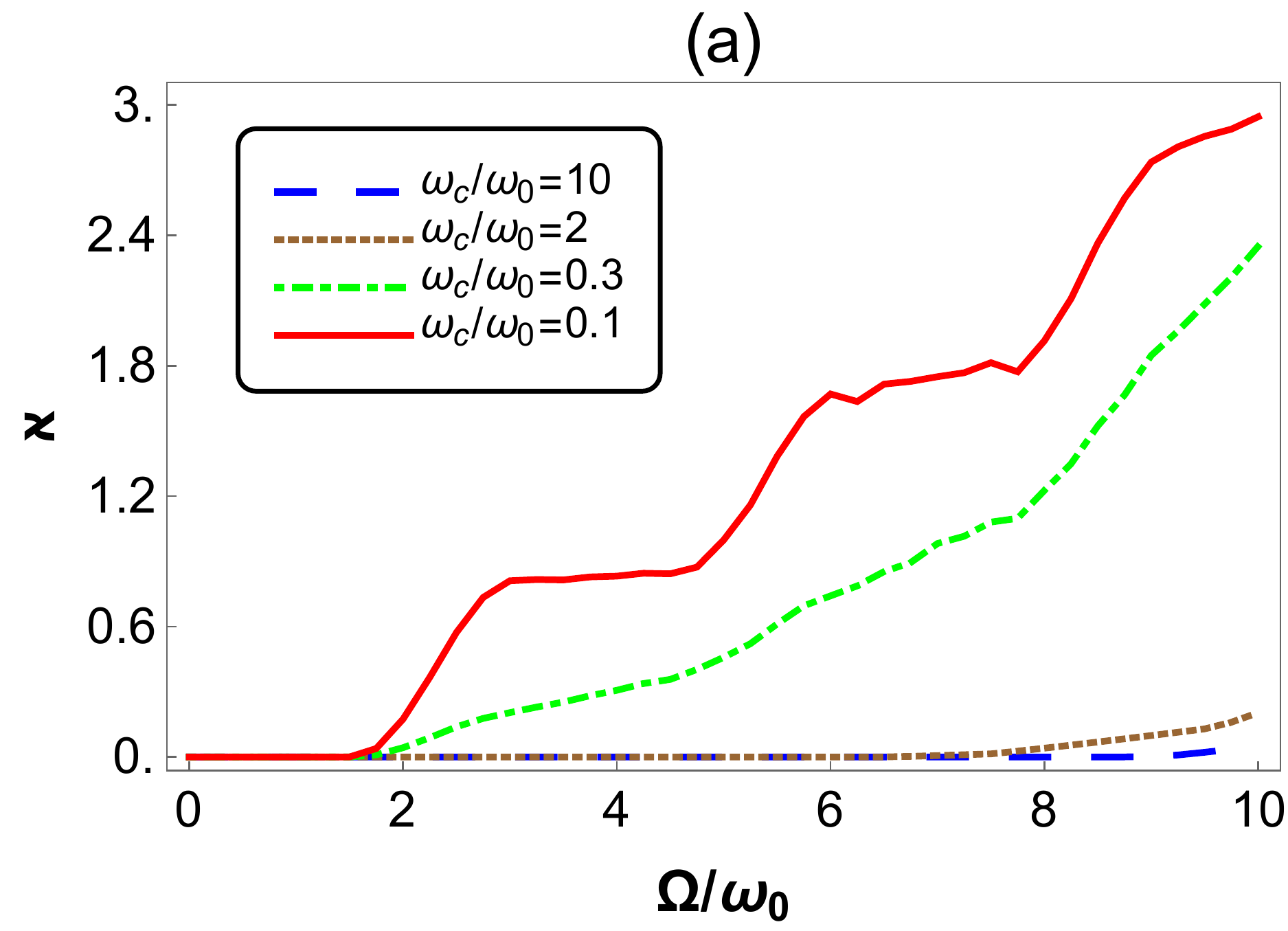}
	\includegraphics[width=5cm,height=4cm]{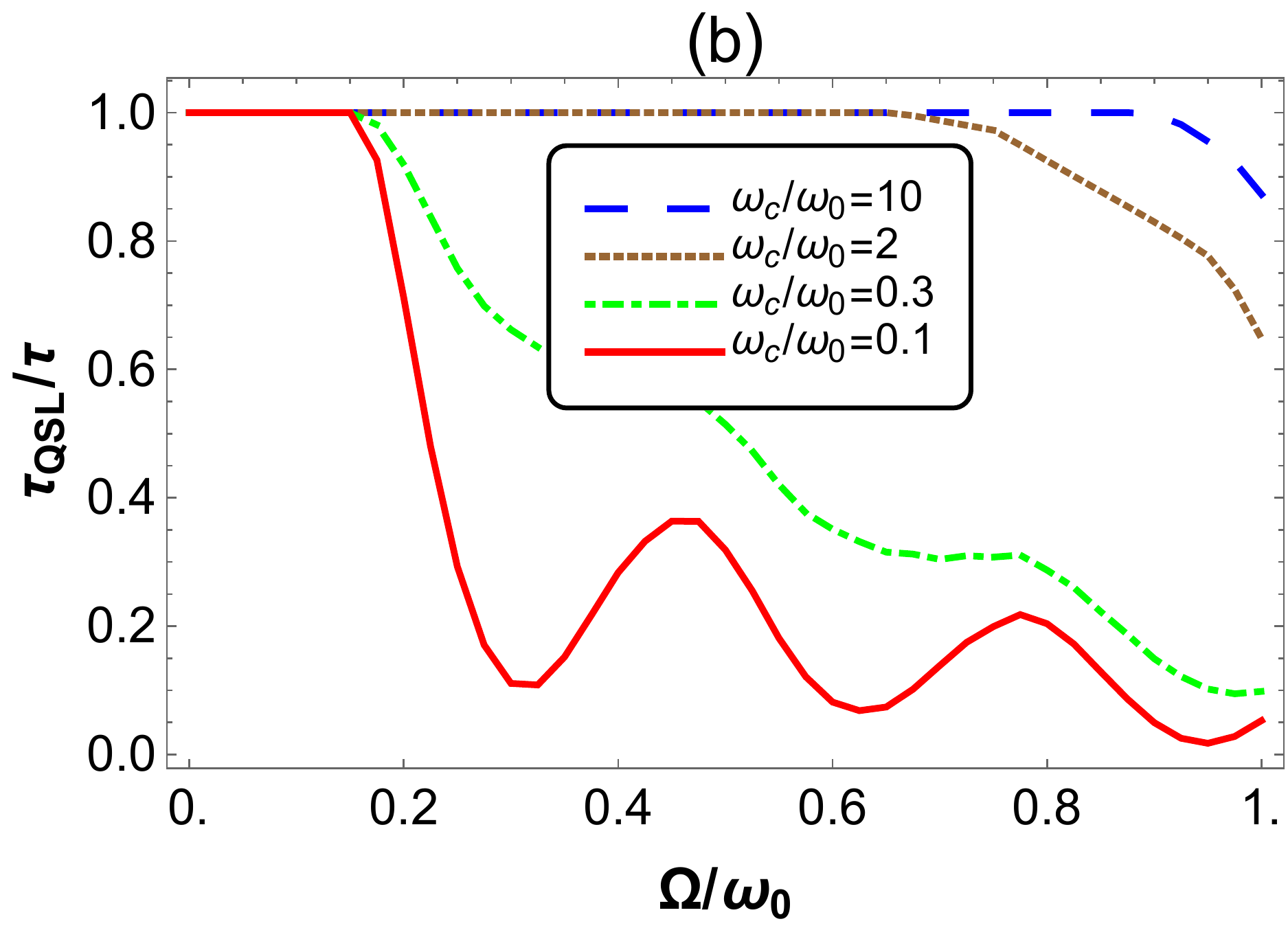}
	\includegraphics[width=5.3cm,height=4cm]{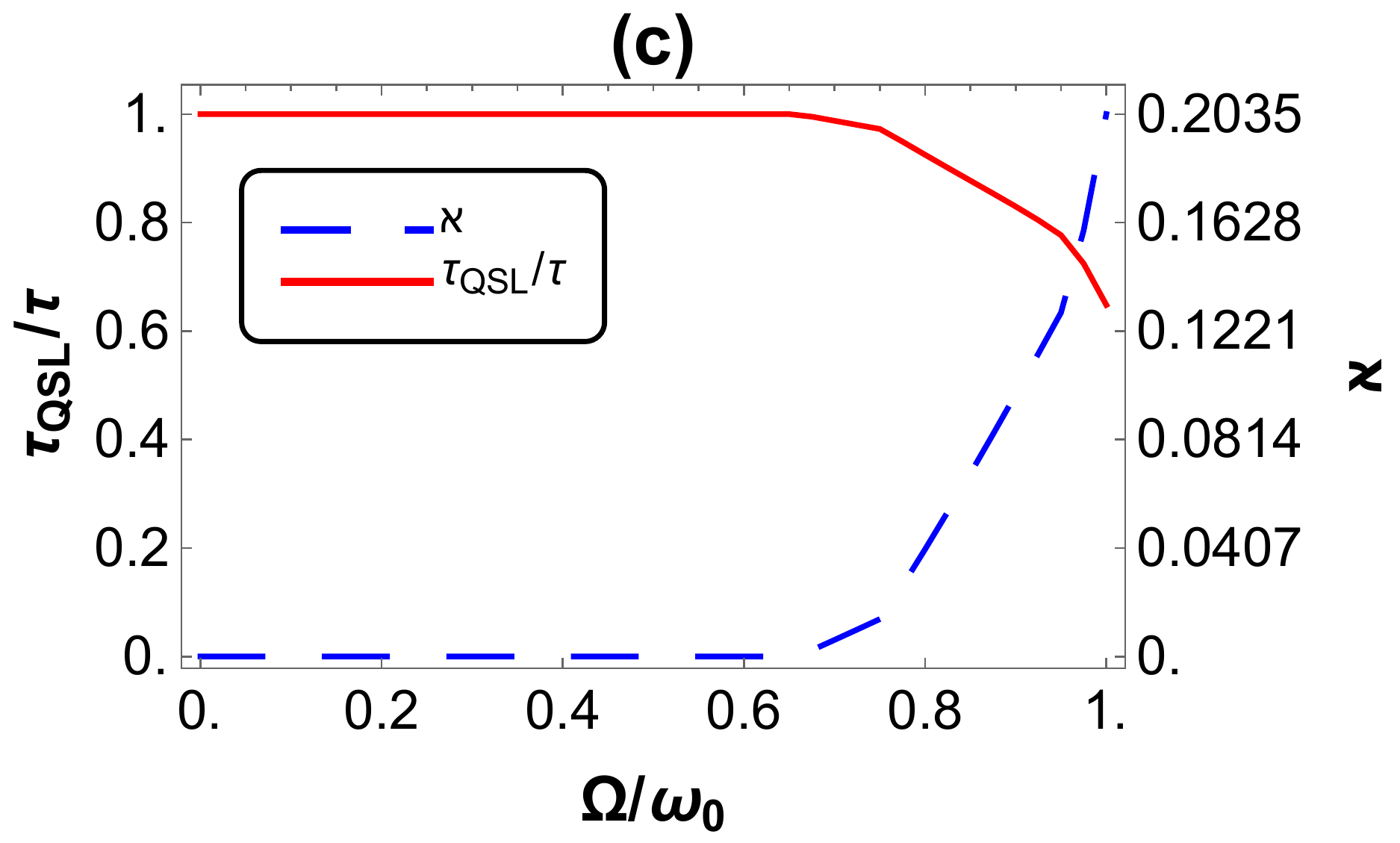}
	\parbox{14.5cm}{\small{\bf Figure 3.}
	(Color online)Influences of the atom-cavity coupling on QSLT and non-Markovianity. (a)non-Markovianity as a function of the coupling strength $\Omega$ in the weak-coupling($\frac{\omega_{c}}{\omega_{0}}=10$, blue dashing line; $\frac{\omega_{c}}{\omega_{0}}=2$, brown dotted line) and strong-coupling($\frac{\omega_{c}}{\omega_{0}}=0.3$, green dot-dashed line; $\frac{\omega_{c}}{\omega_{0}}=0.1$, red solid line) regimes, respectively; (b)QSLT as a function of the coupling strength $\Omega$ in the weak-coupling($\frac{\omega_{c}}{\omega_{0}}=10$, blue dashing line; $\frac{\omega_{c}}{\omega_{0}}=2$, brown dotted line) and strong-coupling($\frac{\omega_{c}}{\omega_{0}}=0.3$, green dot-dashed line;$\frac{\omega_{c}}{\omega_{0}}=0.1$, red solid line) regimes, respectively; (c)QSLT(red solid line) and non-Markovianity(blue dashing line) as a function of the coupling strength $\Omega$ when $\frac{\omega_{c}}{\omega_{0}}=2$.}
\end{center}

We may give the physical interpretation of Figure 3. The non-Markovianity $\mathcal{N}$ is determined by the coupling $\Omega$, the frequency $\omega_{0}$ of the cavity mode and the cut-off frequency $\omega_{c}$, as shown in Eqs.~(\ref{EB306}), ~(\ref{EB402}) and ~(\ref{EB323}). The larger the atom-cavity coupling $\Omega$, the more information the cavity feeds back to the atom, so the non-Markovianity $\mathcal{N}$ will increase with increasing $\Omega$. The information flows irreversibly from the cavity to the reservoir due to the dissipation of reservoir if $\omega_{c}>\omega_{0}$ but the information will flow back from the reservoir to the cavity due to the memory and feedback effect of reservoir if $\omega_{c}<\omega_{0}$, therefore the smaller $\frac{\omega_{c}}{\omega_{0}}$ value causes to the larger non-Markovianity $\mathcal{N}$. Hence the non-Markovianity $\mathcal{N}$ increases and the critical value of sudden transition decreases with decreasing $\frac{\omega_{c}}{\omega_{0}}$, as shown in Figure 3(a). The influence of all parameters on the QSLT may be similarly analyzed, as shown in Figure 3(b), but we omit it in order to save the space. The larger $\mathcal{N}$ induces the smaller QSLT, as shown in Figure 3(c), which is according to Eq.~(\ref{EB403}).

Besides, from Figure 1 and 3 (Figure 2), we also find that the curves of the QSLT and the non-Markovianity vs $\Omega$ ($\delta$) show obvious oscillation behavior, the physical explanation is the exchange of information between the atom and the environment. When the quantum information flows back from the environment to the atom, the non-Markovianity quickly increases and the QSLT quickly decreases. However, when the quantum information flows from the atom to the  environment, the non-Markovianity  will decrease or increase very slowly, and the QSLT will increase.

\section{Conclusions}
In summary, we investigated the QSLT of an atom in a dissipation cavity and the non-Markovianity in the dynamics process. First, we gave an analytical solution (see Eq.~(\ref{EB305})) of the atom and obtain the expressions of the quantum speed limit and the non-Markovianity (see Eq.~(\ref{EB402}) and ~(\ref{EB403})) when the reservoir is at zero temperature and the total excitation number is $N=1$. Furthermore, we demonstrated that, the transition from Markovian to non-Markovian dynamics is the main physical reason of the speed-up process. Both of the atom-cavity coupling and the detuning can induce this transition from Markovian to non-Markovian dynamics. For the Lorentzian reservoir, the sudden transition from no speed-up to speed-up occurs at a same critical point  for different $\lambda$ if $\delta=0$. However, this sudden transitions can occur at different critical points for different $\lambda$ if $\Omega=0.01\gamma_{0}$. The smaller $\lambda$ value causes to the quicker speedup evolution. For the Ohmic reservoir, this sudden transition from no speed-up to speed-up depends on the values of $\frac{\omega_{c}}{\omega_{0}}$. The smaller $\frac{\omega_{c}}{\omega_{0}}$ value causes to the quicker speedup evolution. Recent experiments allow one to drive the open system from Markovian to non-Markovian regimes \cite{Huang,Lyyra}. These results may offer interesting perspectives for future applications of open quantum systems in quantum optical, microwave cavity QED \cite{Varcoe}, trapped ions \cite{Jonathan}, superconducting qubits \cite{Nori} and the Nitrogen-Vacancy center of diamond \cite{Prawer}.
\section*{Acknowledgments}
This work was supported by the National Natural Science Foundation of China (Grant No 11374096) and the Doctoral Science Foundation of Hunan Normal University, China.

\end{document}